# Scaling features of the tribology of polymer brushes of increasing grafting density around the mushroom – to – brush transition

E. Mayoral[1], J. Klapp[1, 2] and A. Gama Goicochea[3†]

[1]Instituto Nacional de Investigaciones Nucleares, Carretera México – Toluca s/n, La Marquesa Ocoyoacac, Estado de México 52750, Mexico

[2]"ABACUS" Centro de Matemáticas Aplicadas y Cómputo de Alto Rendimiento, Departamento de Matemáticas, Centro de Investigación y de Estudios Avanzados (CINVESTAV-IPN), Carretera México-Toluca Km 38.5, La Marquesa, Ocoyoacac, Estado de México, 52740, Mexico

[3]División de Ingeniería Química y Bioquímica, Tecnológico de Estudios Superiores de Ecatepec, Av. Tecnológico s/n, Ecatepec, Estado de México 55210, Mexico

## ABSTRACT

Non equilibrium coarse – grained, dissipative particle dynamics simulations of complex fluids, made up of polymer brushes tethered to planar surfaces immersed in a solvent yield non monotonic behavior of the friction coefficient as a function of the polymer grating density on the substrates, $\Gamma$, while the viscosity shows a monotonically increasing dependence on $\Gamma$. This effect is shown to be independent of the degree of polymerization, *N*, and the size of the system. It arises from the composition and the structure of the first particle layer adjacent to each surface that results from the confinement of the fluid. Whenever such layers are made up of as close a proportion of polymer beads to solvent particles as there are in the fluid, the friction coefficient shows a minimum, while for disparate proportions the friction coefficient grows. At the mushroom to brush transition (MBT) the viscosity scales with an exponent that depends on the characteristic exponent of the MBT (6/5) and the solvent quality exponent ($\nu$ = 0.5, for theta solvent), but it is independent of the polymerization degree (*N*). On the other hand, the friction coefficient at the MBT scales as $\mu \sim N^{6/5}$, while the grafting density at the MBT scales as $\Gamma \sim N^{-6/5}$ when friction is minimal, in agreement with previous scaling theories. We argue these aspects are the result of cooperative phenomena that have important implications for the understanding of biological brushes and the design of microfluidics devices, among other applications of current academic and industrial interest.

† Corresponding author. Electronic mail: agama@alumni.stanford.edu

## I INTRODUCTION

Friction, lubrication and wear at the atomic level are known to be important to understand the behavior of, and help in the design of new nanomaterials for specific applications [1]. It is an empirical fact that a smooth, homogeneous surface shows little friction. In the context of complex fluids, it has been shown that the friction coefficient ($\mu$) of a fluid confined by surfaces can be reduced by up to three orders of magnitude when polymer brushes are attached to the substrates [2]. This problem is of considerable interest not only because of its obvious applications to lubricants and in nanotribology [1, 3-5], but also because there are biological examples of it in synovial joints, drug – delivering liposomes, and other similar systems. Biocompatible polymer coatings soluble in water, such as poly (ethylene glycol) (PEG), are used in many applications [6]. As an example, liposomes containing surface-grafted PEG resist adsorption of diverse components; this feature makes it possible to design sterically stabilized liposomes appropriate for drug delivery.

Although several works have dealt with rheological studies of polymer brushes, few if any deal with the dependence of the friction coefficient with grafting density [7, 8]. Yet, this is a problem of relevance because varying the density of polymers grafted to a surface is equivalent to controlling the mushroom to brush transition (MBT), which can be used to design stimuli – responsive materials [9, 10]. When polymers are grafted at low densities they form structures that resemble mushrooms, whereas when there are many of them on a surface they tend to form aligned structures, resembling a brush. Numerical tribology studies of fluids confined by surfaces under flow have shown that the effective friction gets reduced substantially when polymer brushes are grafted to these surfaces [11]. The influence of shear rate, solvent quality and ionic strength on the friction coefficient of polyelectrolyte brushes were studied in [12], while the effect of the compression rate of polymer and polyelectrolyte brushes of fixed grafting density was reported in [13]. It was found that the friction coefficient correlates directly with the degree of interpenetration of brushes, being larger for neutral polymer brushes than for polyelectrolyte brushes [12]. On the other hand, brushes made of polymers or polyelectrolytes were found to yield approximately the same coefficient of friction when compressed by the same amount [13]. For a comprehensive and recent review about lubrication between polymer brushes the reader is referred to [14]. Evidently, this is a feature that has enormous potential for applications in lubricants, but there are also important aspects of basic science that require elucidation, such as the dependence of the viscosity of a complex fluid on the concentration of the polymers grafted to the confining surfaces. The MBT has not yet been monitored as a function of the changes in measurable properties such as the friction coefficient and the viscosity of the fluid, and that is the aim of the present work.

It is well known that controlled applications in nanotribology can be developed using grafted polymer brushes varying two parameters: the polymerization degree of the chains, $N$, and the polymer grafting density $\Gamma = N_p/A$, where $N_p$ is the number of polymer chains grafted on a surface of area $A$. These parameters define the MBT. At low chain grafting density, the mushroom regime is present and increasing such density the brush regime appears. Many properties differ significantly between the mushroom and brush regimes. In the mushroom regime, small concentrations lead to essentially no interaction between chains, and the they adopt random configurations with characteristic dimension given by the Flory radius, $R_F$, depending on $N$ and the size of the monomer unit, $a_m$, similarly to free chains in solution. In this case, the length of the free polymer is given by excluded volume effects, $v_m =$ $a_m^3(1-2\chi)$, which produce an increase in the size of the chain, *R*. Here, *R* is the end-to-end distance of the chain. In an athermal solvent, the Flory – Huggins interaction parameter $\chi$ = 0 (intramolecular interactions could be ignored) and $v_m \approx a_m^3$. The three - dimensional Flory radius of a chain with excluded volume interactions is [15]

$$R_{F3} \approx a_m N^{\nu} , \tag{1}$$

where $\nu$ = 3/(*d* +2) and *d* being the spatial dimension, which fairs reasonable well when compared with experiments and numerical calculations [16, 17]. The scaling exponent $\nu$ is known to depend on solvent quality also [15].

As soon as the grafting density is increased, there appears a concentration at which polymer head groups start to interact with one another, $\Gamma_{MBT}$, adopting a more stretched configuration, resembling a brush. This transition displays a scaling law for $\Gamma_{MBT}$ with *N*, which, is given by [18]:

$$\Gamma_{MBT} \propto \left(\frac{A}{\pi a_m^2}\right) N^{-6/5} \quad . \tag{2}$$

The main feature in the brush regime is the one – dimensional nature of the polymer confined in this region. The thickness of the polymer brush and their free energy are linear functions of *N* [18].

The flow of entangled polymers grafted on surfaces has special characteristics. Their slippage is described by the distance to the wall at which the velocity extrapolates to zero, known as the extrapolation length, *b*. The coefficient of friction $\mu$ is defined as the ratio between the mean forces that the grafted beads on the surfaces experience along the flow direction, $\langle F_x(\dot{\gamma})\rangle$, and perpendicularly to it, $\langle F_z(\dot{\gamma})\rangle$, namely:

$$\mu = \frac{\langle F_x(\acute{\gamma})\rangle}{\langle F_z(\acute{\gamma})\rangle} . \tag{3}$$

The shear stress ($\sigma_x$) is related with the viscosity, $\eta$, as follows:

$$\sigma_x = \frac{\langle F_x(\acute{\gamma})\rangle}{A} = \eta\dot{\gamma} \quad . \tag{4}$$

In eq. (4), $\acute{\gamma}$ is the shear rate, which is the velocity gradient in the pore formed by opposing surfaces under stationary, Couette flow [19]. The extrapolation length $b$ can be obtained as the ratio between the viscosity $\eta$ and the friction coefficient $\mu$: $b = \eta/\mu$. For semi-ideal conditions [20], i.e., in the case where a small amount of chains with large $N$ are grafted on the wall (mushroom regime) a linear dependence of $\mu$ with $\Gamma$ is found [20]

$$\frac{\sigma_x}{V} = \mu \approx \ [\Gamma\eta R_{F3}] , \tag{5}$$

where $V$ is the velocity of the stationary flow. Then, the extrapolation length can be written as [20]:

$$b = (\Gamma R_{F3})^{-1}. \tag{6}$$

The normal stress, $\sigma_z$, also obeys scaling laws. A general scaling form was presented by Alexander and de Gennes for the osmotic pressure between parallel plates covered with polymer brushes of polymerization degree $N$ separated by a distance $D$ [21]:

$$\sigma_z = \ (k_B T)\ f(a_m, D, N)\Gamma^y , \tag{7}$$

where $f$ is a function that does not depend on the grafting density $\Gamma$. The scaling exponent of the grafting density in equation (7) is defined in terms of the scaling exponent ($\nu$) of the Flory radius $R_{F3}$: $y = 3\nu/(3\nu - 1)$. Recent numerical simulation studies [8] have shown that eq.

(7) is fulfilled for polymer brushes of increasing grafting density under Couette flow immersed in theta solvent, where $\nu$ = ½ and $y$ = 3.

In this work, we carry out mesoscopic scale simulations of linear polymer chains grafted on two parallel surfaces under stationary flow and calculate their viscosity and friction coefficient as functions of increasing grafting density. In Section II we present the models and methods used in this work, as well as all the details pertaining the simulations performed. The results obtained and their discussion are to be found in Section III, followed by our conclusions, in Section IV.

## II MODELS AND METHODS

We have performed dissipative particle dynamics (DPD) simulations of linear grafted polymers immersed in a solvent, in the canonical ensemble (fixed density and temperature), under stationary, Couette flow. The DPD model is by now well-known [22 – 25], therefore we shall reproduce only what is pertinent here. Three forces make up the basic DPD model: a conservative force ($\vec{F}_{ij}^{C}$), that accounts for the local pressure of the fluid and is proportional to the interaction constant $a_{ij}$; a dissipative force ($\vec{F}_{ij}^{D}$), which represents the viscosity arising from collisions between particles, proportional to the (negative) relative velocity of the particles and to a constant, $\gamma$; and a random force ($\vec{F}_{ij}^{R}$), that models the Brownian motion of the particles, with an intensity given by the constant $\sigma$ (not to be confused with the shear or normal stresses). These forces are all short ranged; in particular, the conservative force is linearly decaying, $\vec{F}_{ij}^{C} = a_{ij}\left(1 - r_{ij}/R_c\right)\hat{e}_{ij}$, where $r_{ij} = r_i - r_j$ represents the relative position vector between particles $i$ and $j$, $\hat{e}_{ij}$ is the unit vector in the direction of $r_{ij}$. The constants in the dissipation and random forces are not independent, and satisfy the relation [25] $\sigma^2/2\gamma = k_B T$, which is the expression for the fluctuation – dissipation theorem in DPD. $R_c$ is the cut off distance, beyond which all forces are zero. The DPD beads are all of the same size, with radius $R_c$, which is set equal to 1.

We obtain the friction coefficient ($\mu$) using eq. (3), and the viscosity ($\eta$) of the fluid, through the relation [26] $\eta = \langle F_x(\acute{\gamma})\rangle / A\,\acute{\gamma}$, see eq. (4), where $\langle F_x(\acute{\gamma})\rangle$ and $\langle F_z(\acute{\gamma})\rangle$ are the mean forces that the particles on the surfaces experience along the flow direction, and perpendicularly to it, respectively; the brackets indicate an ensemble or time average. Those forces are obtained from the components of the pressure tensor, which in turn are obtained from the virial theorem. The shear rate $\acute{\gamma}$ is defined as $2v_0/D$, where $v_0$ is the constant flow velocity exerted on the wall with grafted particles, and $D$ is the separation between the opposite surfaces; both of these parameters are kept constant in this work, see Fig. 1. The extrapolation length $b$, defined before as $b = \eta/\mu$ is calculated using eq. (8):

$$b = \frac{\langle F_x(\acute{\gamma})\rangle / A}{\acute{\gamma}} \ . \tag{8}$$

We modeled brushes made of linear homopolymers of various polymerization degrees, where only the "head" of each chain is grafted on the surface. Two sizes of the parallelepiped simulation box were constructed to test finite size effects, one relatively small and one large, with sides on the square *xy* – plane equal to $L = 7$ and $L = 50$, respectively, in reduced DPD units. The length of the box in the *z* – direction was set at $D = 7$ and $D = 5$ for the small and large boxes, respectively. The total number of DPD beads was fixed to yield a total number density $\rho = 3$ in all cases, regardless the grafting density or the polymerization degree of the chains. For this reason when the grafting density $\Gamma = N_p/A$ increases, the number of solvent molecules must be reduced. $N_p$ is the number of chains of polymerization degree $N$ tethered on each wall of area $A$; $N_m = NN_p$ is the total number of monomeric units making up the chains in the system and $N_T = N_s+N_m$ is the total number of DPD units in the simulation box ($N_s$ is the number of solvent beads). For later purposes, it is convenient to define the fraction $\phi$ as follows:

$$\phi = \frac{NN_p}{N_s} = \frac{N_m}{N_s} \ . \tag{9}$$

The fraction defined by eq. (9) is important because it is helpful to signal the MBT transition, when $\phi \approx 1$, as we shall see later. The conservative force interaction parameter, $a_{ij}$, was chosen equal to 78 units for particles of the same type ($i = j$), as well as for solvent – monomer interactions; this choice defines theta – solvent conditions. We have chosen to model brushes under theta conditions so that the results are not dependent on the choice of interaction parameters $a_{ij}$. The surfaces on which the chains were grafted are effective walls modeled with the force $F_{wall}(z) = a_{wi}[1 - z_i/z_C]$, introduced for the first time by one of us [27], whose direction is perpendicular to the $xy$ – plane, with wall interaction constant $a_{wi} = 70$ for the polymers' head grafted on the surface, and $a_{wi} = 100$ otherwise (the rest of the monomers in the chains, and the solvent beads). $F_{wall}$ becomes zero when the distance of the $i$-th particle from the wall along the $z$ – axis, $z_i$, becomes larger than $z_C = 1$. To implement conditions of Couette flow a constant velocity along the $x$ – axis is imparted to the grafted heads of the chains only, on that surface, which is the shear velocity, $v_0$, see Fig. 1. The velocity imparted to the polymers' heads on the opposite surface is $-v_0$, and it was fixed at $v_0$ = 1.0; the temperature was chosen as $k_BT = 1$; the finite time step for the integration of the equation of motion was $\delta t = 0.01$; all quantities used in this work are expressed in reduced DPD units. Our results were obtained from averages of simulations of up to $4\times10^3$ blocks, of $2{\times}10^4$ time steps each, using first $2{\times}10^3$ blocks for equilibration and the rest for the production phase; when properly dimensionalized this represents a time observation window of 0.16 ms.

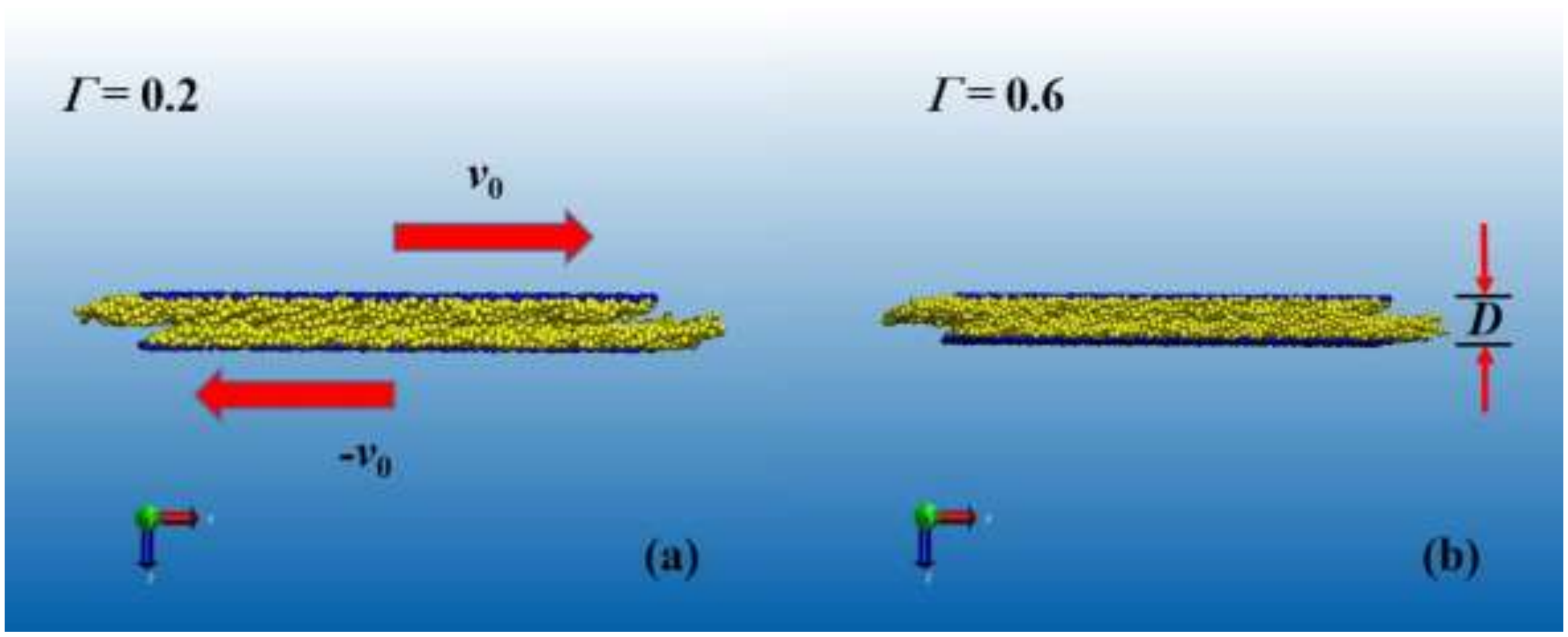

**Figure 1.** (Color online) Snapshots for two different grafting densities of the fluid made up of polymer brushes: **(a)** $\Gamma$ = 0.2 and **(b)** $\Gamma$ = 0.6 with $N$=10; the solvent is removed for clarity. The heads of the polymers grafted to the membranes are shown in blue, the rest of the brush chains are in yellow. For both cases, $L_x = L_y =$ 50 and $D$ = 5, and the shear rate is $\dot{\gamma} = 0.028$, in reduced DPD units. The shear velocity $v_0$ imposed on the (blue) grafted beads on each wall is also indicated.

All simulations were carried out at constant particle number, volume of the simulation box, and temperature, i.e. using the canonical ensemble. Confined systems are usually studied at constant chemical potential, volume and temperature (grand canonical ensemble); however, as shown by Goujon and collaborators [28], the rheology of polymer brushes is not affected by the choice of ensemble once the stationary flow is established. This is a useful result because grand canonical simulations are considerably more time consuming than their canonical counterparts.

## III RESULTS AND DISCUSSION

Let us start by presenting the results corresponding to the smaller simulation box, which is a cube of side $L = 7$. The values of the polymerization degree modeled were $N$ = 7, 8, 10, 12, 17, and 21. As shown in Fig. 2, the friction coefficient decreases with increasing $\Gamma$ until reaching a minimum, $\Gamma_{\min}$, and then it rises again. This highly nonlinear behavior for $\mu(\Gamma)$ is accompanied by monotonous increase in the viscosity $\eta$ when the grafting density $\Gamma$ grows, as is commonly observed in other systems [8]. The same trends are found for the larger system whose volume is $V = 50 \times 50 \times 7$, presented in Fig. 3, which confirms that finite size effects are not predominant in these cases [29].

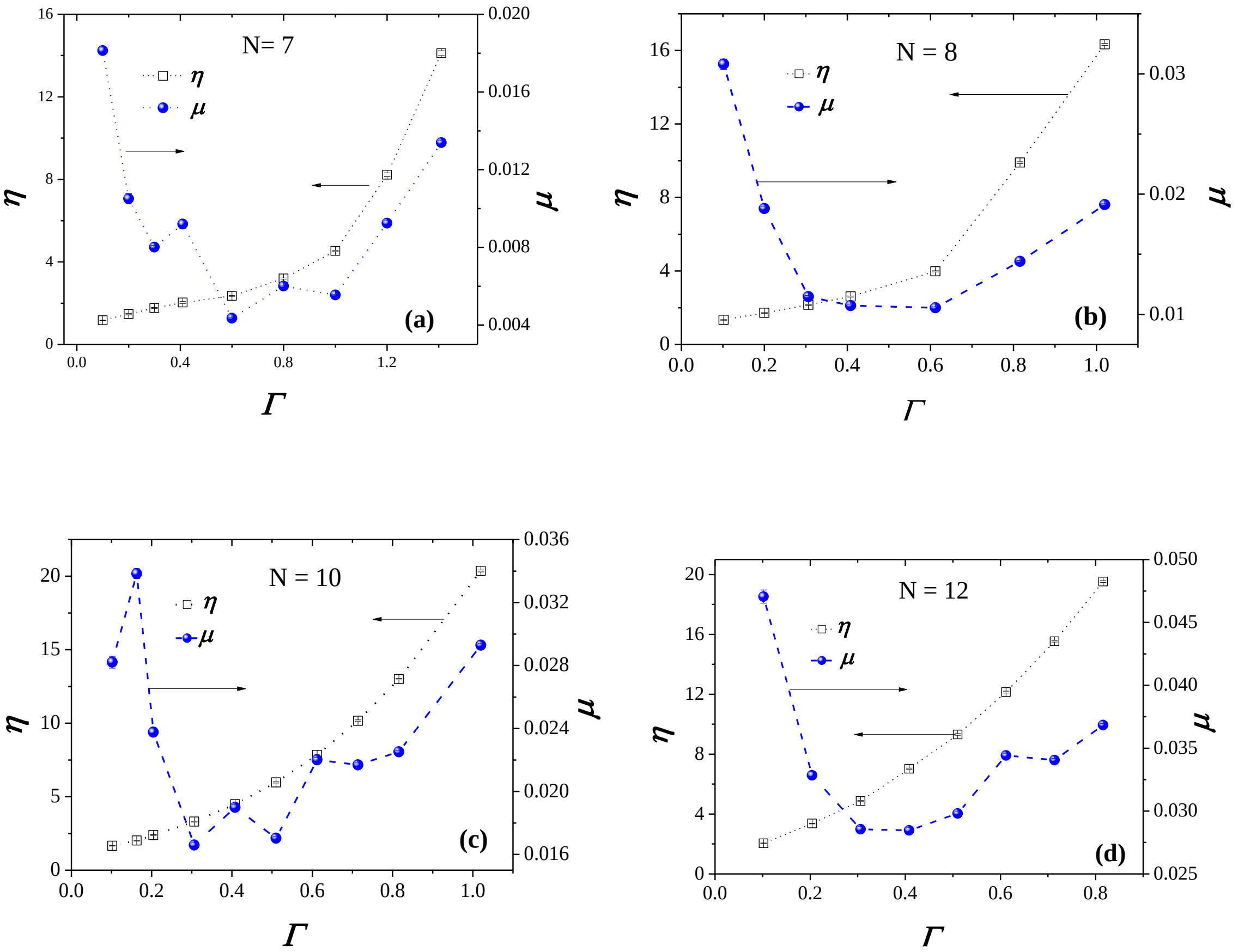

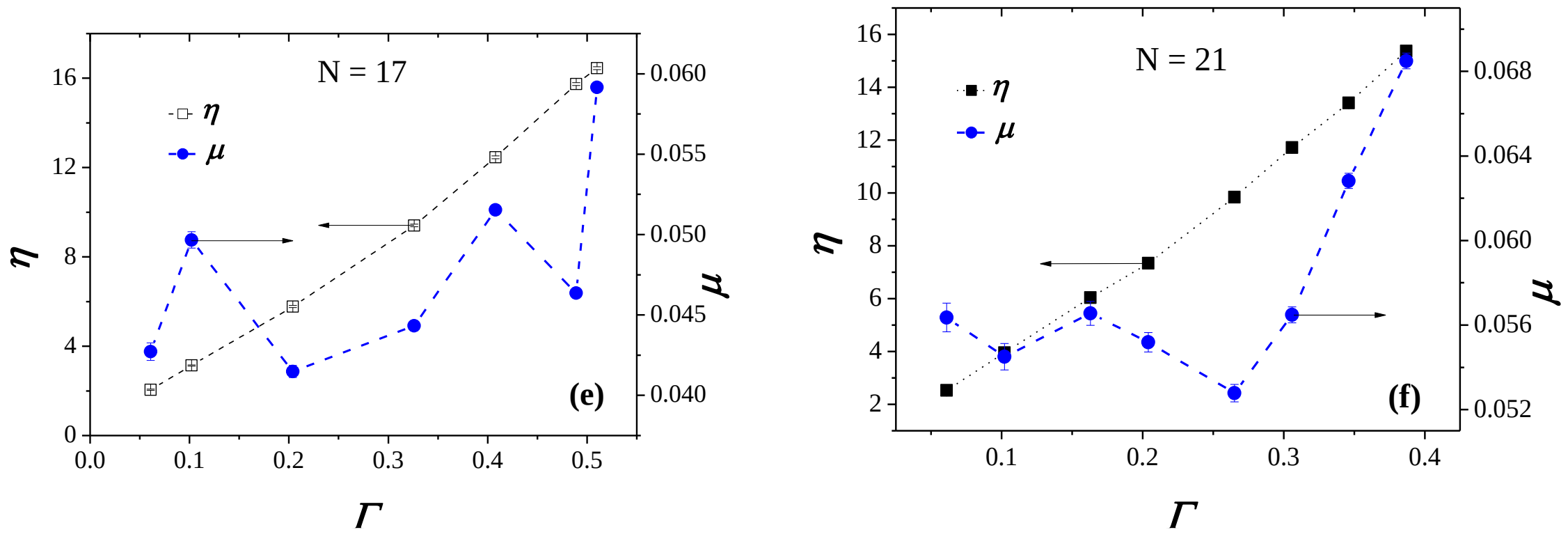

**Figure 2**. Viscosity ($\eta$, squares) and friction coefficient ($\mu$, solid circles) of polymer brushes as functions of the grafting density ($\Gamma$), for different values of the degree of polymerization of the brushes: (a) $N$ = 7, (b) $N$ = 8, (c) $N$ = 10, (d) $N$ = 12, (e) $N$ = 17, and (f) $N$ = 21. In all cases, $L_x = L_y = D = 7$, and the shear rate is $\dot{\gamma} = 0.028$. Error bars are smaller than the symbols' size; dashed lines are only guides for the eye.

The non – monotonous behavior displayed by the friction coefficient is found to be independent of the system's size, which is explained as due to the fact that the surfaces can be considered as being rough, that is self – similar, on account of them being covered by a non – uniform layer of chains. The entanglement between the chains on opposite surfaces is modified by the amount of solvent beads and for this reason the brushes can slide over one another, for specific grafting densities. This effect arises from the composition of the first particle layer adjacent to each surface, resulting from the confinement of the fluid.

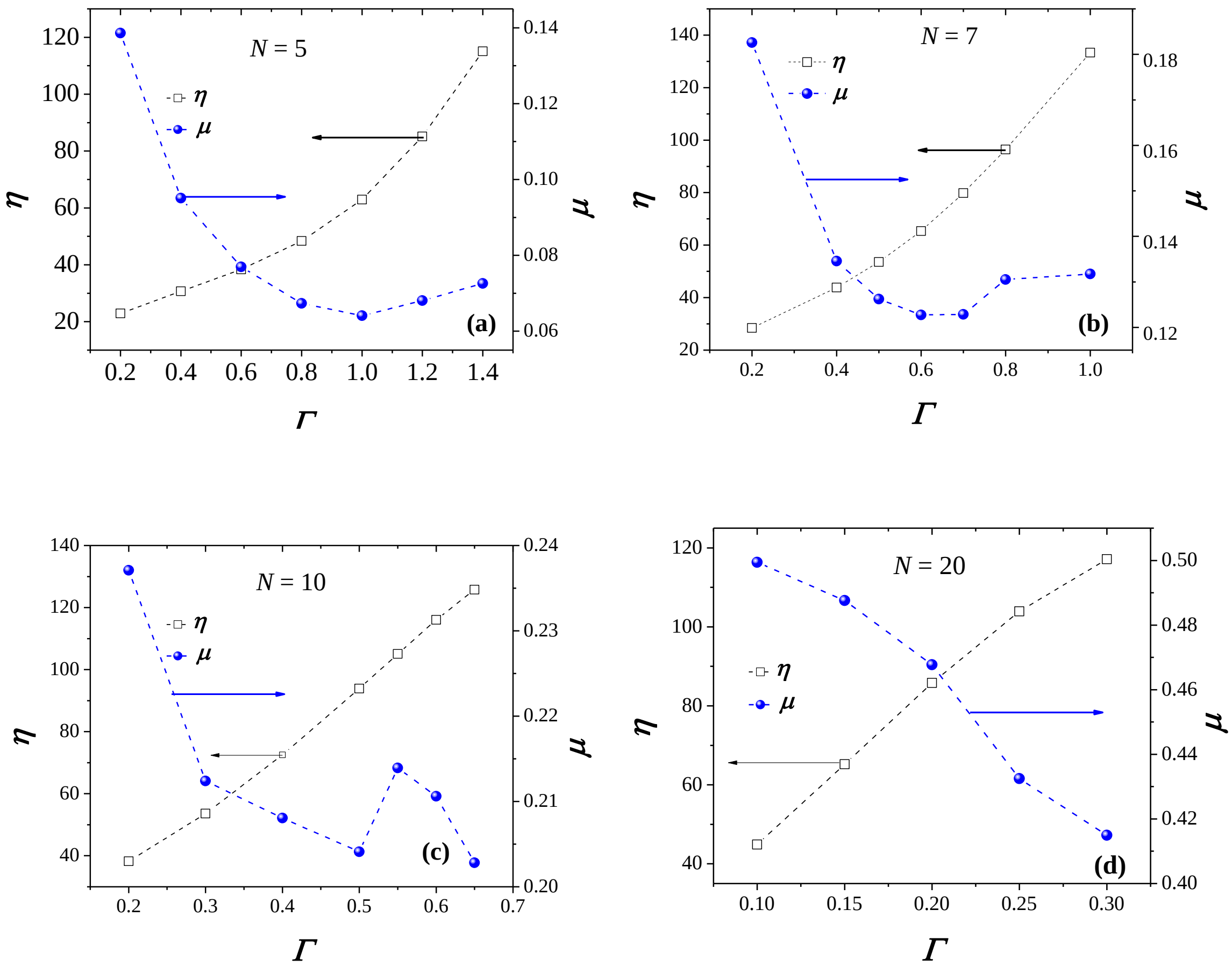

**Figure 3**. Viscosity ($\eta$, empty squares) and friction coefficient ($\mu$, solid circles) of polymer brushes as functions of the grafting density ($\Gamma$), for different values of the degree of polymerization of the brushes: (a) $N$ = 5, (b) $N$ = 7, (c) $N$ = 10, (d) $N$ = 20. In all cases, $L_x$=$L_y$= 50 and $L_z$ = $D$ = 7, and the shear rate is $\dot{\gamma} = 0.028$. All quantities are reported in reduced units. Error bars are smaller than the symbols' size; dashed lines are only guides for the eye.

The density profiles for both solvent and brush for the larger system, presented in Fig. 4 for $N$ = 10, show how each component is distributed within the pore formed by the parallel surfaces. The density profiles for all values of the polymerization degree modeled in this work can be found in Fig. A1, in the Appendix, which are omitted here for brevity. At low grafting densities (see Fig. 4) the solvent beads penetrate the sparsely grafted chains and there is an extensive chain-solvent interface. When the grafting density is increased, a gradual increase of the polymer layering at the solid substrate is obtained and the solvent is excluded toward the bulk. The brush-solvent interaction reduces the entanglement between the chains and as a consequence the friction coefficient is reduced. This trend continues with increasing $\Gamma$ until some value of $\Gamma_{min}$ is reached, when the chains become a brush and can be considered as an attractive solid surface, leading to the increase in friction as the grafting density grows.

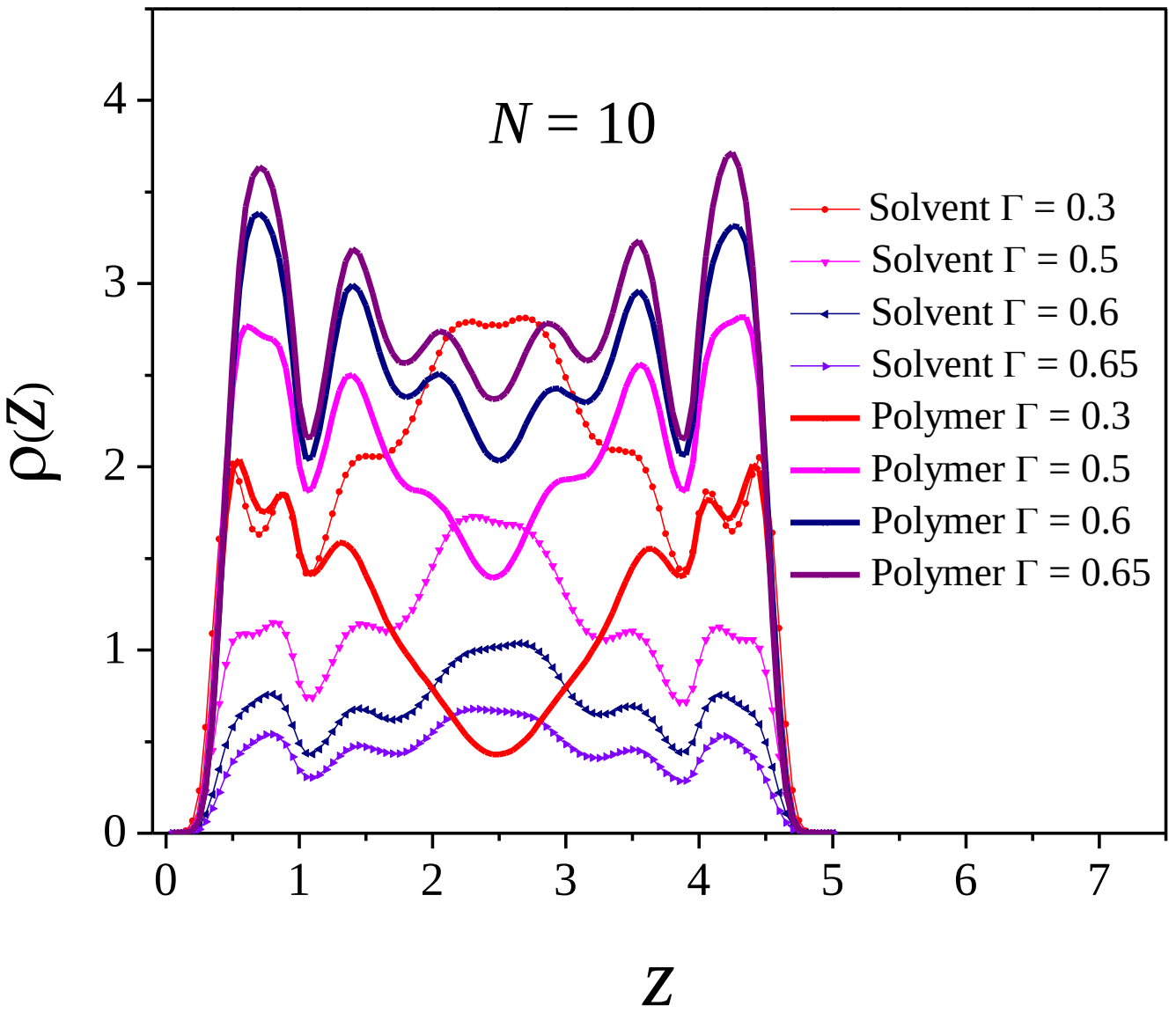

**Figure 4**. Density profiles for the solvent (symbols) and brush monomers (lines) of the larger system at increasing values of the grafting density $\Gamma$, for polymerization degree equal to $N = 10$. The density profiles for all polymerization degrees modeled can be found in Fig. A1, in the Appendix. All quantities are reported in DPD units.

In the high-grafting regime, slippage is expected to occur because the brush roughness is decreased, but in our case this is accompanied by the reduction of solvent beads to keep the global density constant, which produces an increase in $\mu$. This stresses the importance of the collective effects between the solvent and the polymer brush. Whenever the particle layers on the surfaces are made up of as close a proportion of polymer beads ($N_m = N\,N_p$) as solvent particles ($N_s$), the friction coefficient shows a minimum, while for disparate proportions the friction coefficient grows. Using the ratio $\phi$ defined in eq. (9), which is the proportion between the number of monomeric units that make up the grafted chains $N_m$ and the solvent particles $N_s$, one finds scaling behavior in the viscosity. Figure 5 shows the dependence of $\eta$ with $\phi$.

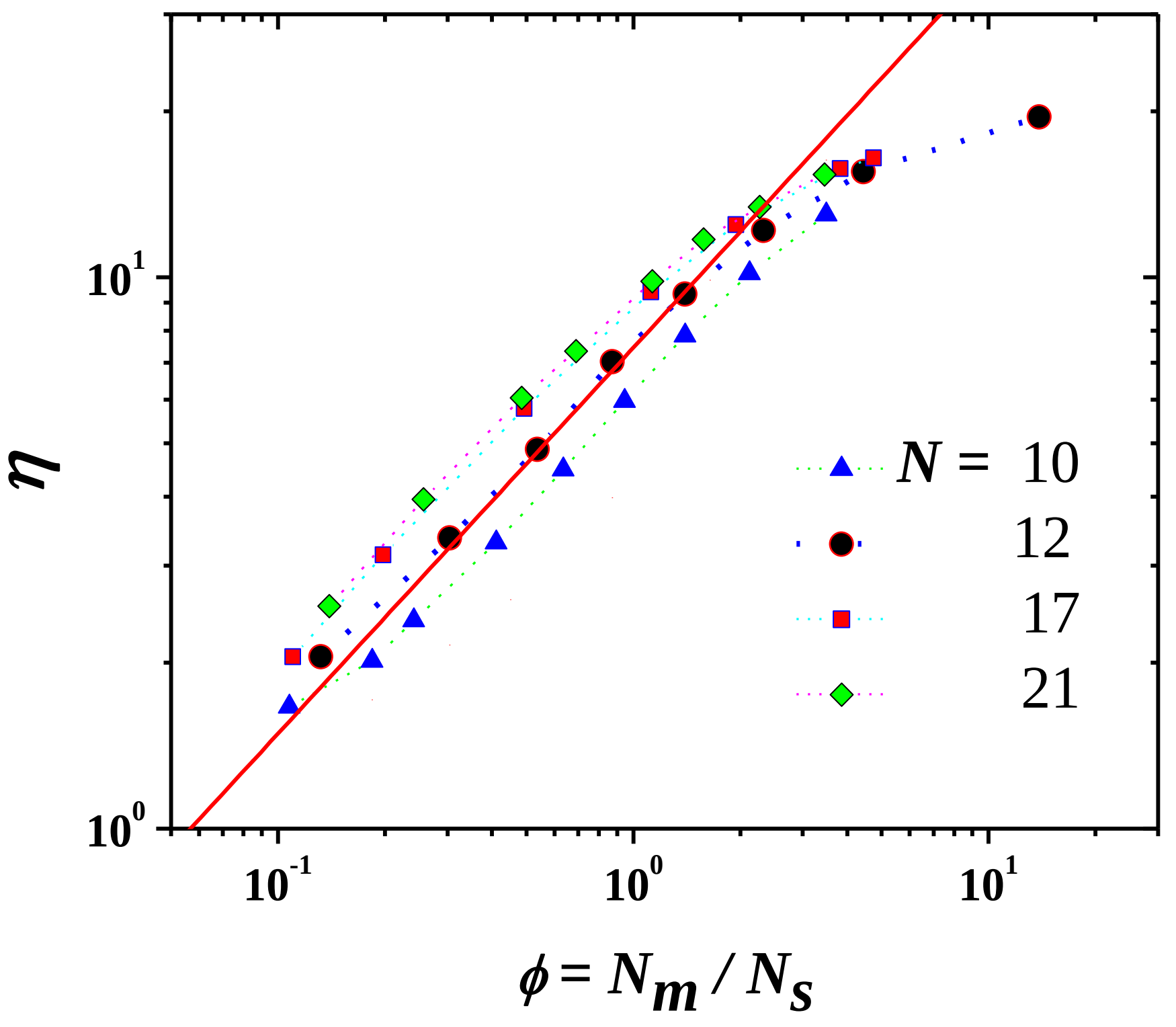

**Figure 5**. Viscosity $\eta$ as a function of $\phi = N_m/N_s$ for four values of the polymerization degree, $N$. The solid line represents the fit $\eta \sim \phi^{\alpha}$, where the exponent $\alpha = 0.70$.

The data for the viscosity presented in Fig. 5 for brushes made up of chains of four different values of the polymerization degree show the same scaling behavior with respect to the ratio $\phi$, defined in eq. (9), namely

$$\eta \sim \phi^{\alpha}, \tag{10}$$

where the scaling exponent $\alpha = 0.7$ is the same for all cases, especially in the neighborhood of $\phi \sim 1$ (which defines the MBT), regardless the polymerization degree. This exponent depends not only on the properties of MBT, but on the solvent quality also, since $\alpha = 6/5 - \nu$, where the first term arises from the MBT – see eq. (2) – and the second is the scaling exponent of the gyration radius, see eq. (1). The latter is 0.5 for polymers in theta solvent, hence $\alpha = 0.7$. We shall have more to say about the exponent $\alpha$ in eq. (10) when the scaling of the friction coefficient is presented. Figure 6(a) shows the behavior of the extrapolation length $b = \eta/\mu$ as a function of $\Gamma$ for $N$=10; for results on all values of $N$, see Fig. A2 in the Appendix. The slope of the linear fit seen in Fig. 6(a) is not universal and depends on the polymerization degree. The trend seen in Fig. 6(a) can be summarized as $b \sim S\Gamma$, where $S$ is the slope of the linear fits. The value of the slope of the linear fits shown in Figs. 6(a) and A2 (in the Appendix) can be used to determine the behavior of $b$ as a function of the polymerization degree. The resulting curve is presented in Fig. 6(b).

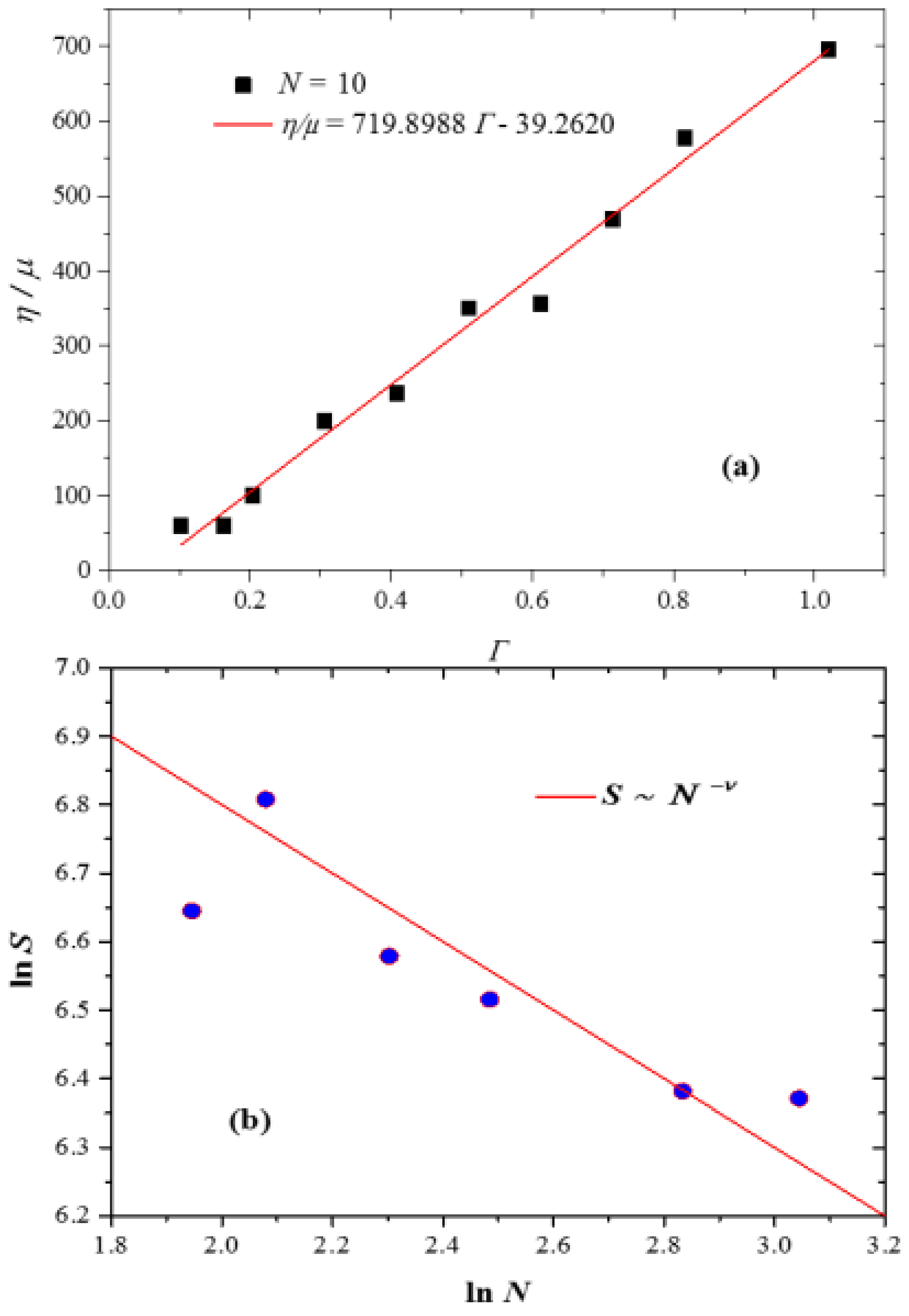

**Figure 6. (a)** The extrapolation length $b$ as a function of the grafting density $\Gamma$ for $N$=10. The line is the best linear fit. The results for all values of $N$ modeled can be found in Fig. A2 in the Appendix. All quantities are reported in reduced units. **(b)** Slope $S$ (symbols) of the linear fits of $b$ vs $\Gamma$, shown in Fig. 6(a), as a function of $N$. The solid line is the fit $S \sim N^{-\nu}$, with $\nu = 0.5$.

As shown in Fig. 6(b), the slope of the extrapolation length obeys a scaling law that can be expressed as $S{\sim}N^{-\nu}$, where the value of the exponent $\nu$ obtained ($\nu = 0.5$) is the one corresponding to the theta – solvent exponent for the Flory radius, ν = ½ [15], which corresponds precisely to the conditions of the solvent we defined when setting up the DPD conservative interaction constants, see the discussion following eq. (9) .

Universal behavior for the extrapolation length can be obtained when it is plotted as a function of $\phi$, see eq. (9), for different values of $N$; the results are summarized in Fig. 7(a).

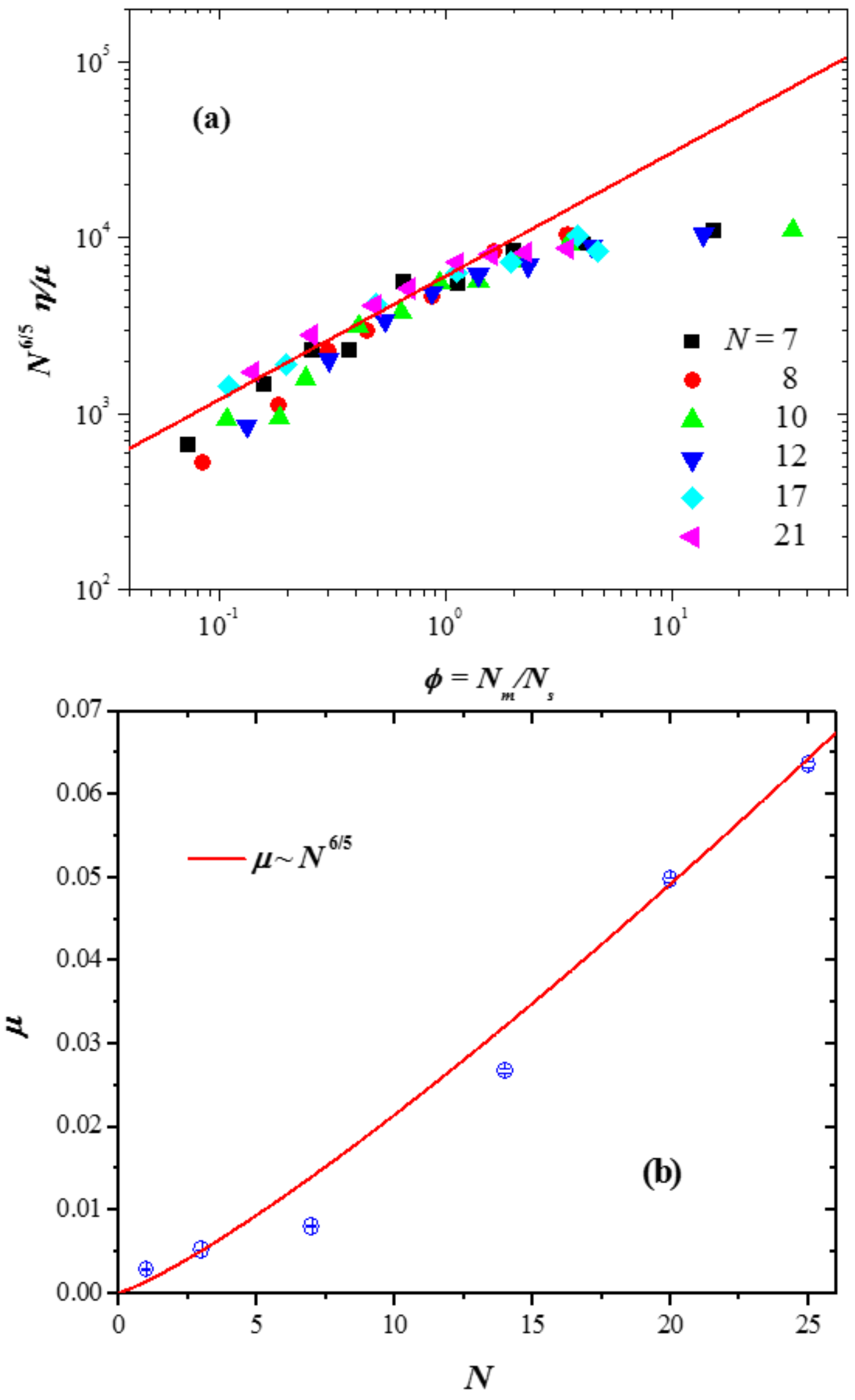

**Figure 7. (a)** Scaling behavior for the extrapolation length $b$, multiplied by the polymerization degree to the 6/5 power, as a function of $\phi$ for different values of *N*. All quantities are reported in reduced units. The solid line is the fit $N^{6/5}b \sim \phi^{\alpha}$, with $\alpha = 0.7$, as in Fig. 5. **(b)** Friction coefficient ($\mu$) of polymer brushes as a function of the degree of polymerization (*N*). In all cases, $L_x$=$L_y$=$D$=7, the shear rate is $\dot{\gamma} = 0.028$, and the grafting density is $\Gamma$=0.30. The line is the fit $\mu{\sim}N^{6/5}$.

The results presented in Fig. 7(a) indicate that $N^{6/5}b$ scales as $\phi^{\alpha}$, with $\alpha = 0.7$, except when the number of monomers in polymer chains is notably larger than the number of solvent particles ($\phi > 1$), which is beyond the MBT and belongs to the brush regime. The scaling exponent is the same as that in the scaling of the viscosity as a function of $\phi$, see Fig. 5. Since $b = \eta/\mu$ and $\eta{\sim}\phi^{\alpha}$, it follows from Fig. 7(a) that the friction coefficient must scale as $\mu{\sim}N^{6/5}$ in the MBT. This scaling is confirmed by the results shown in Fig. 7(b), where the friction coefficient is obtained as a function of the polymerization degree at a fixed grafting density.

The $N^{6/5}$factor in Fig. 7(a) (and in the scaling found in Fig. 7(b)) has its origin in the MBT, see eq. (2), and it is discussed in more detail now. From Figs. 2 and 3, there appears a minimum value of the friction coefficient that varies with the chain grafting density; such minimum occurs at the MBT, $\Gamma_{MBT}$, in each system. The value of $\Gamma$ at the minimum in $\mu$, $\Gamma_{min}$ = $\Gamma_{MBT}$, is shown as a function of the degree of polymerization *N* in Fig. 8. A power law dependence is obtained, $\Gamma{\sim}N^{\omega}$, where the exponent is equal to $\omega$ = - 6/5 for the MBT, as expected from eq. (2). This confirms that the first minimum of $\mu$ corresponds to the MBT. Also, this transition is responsible for the universal behavior in $b$ as is suggested by the rescaling factor, shown in Fig. 7(a).

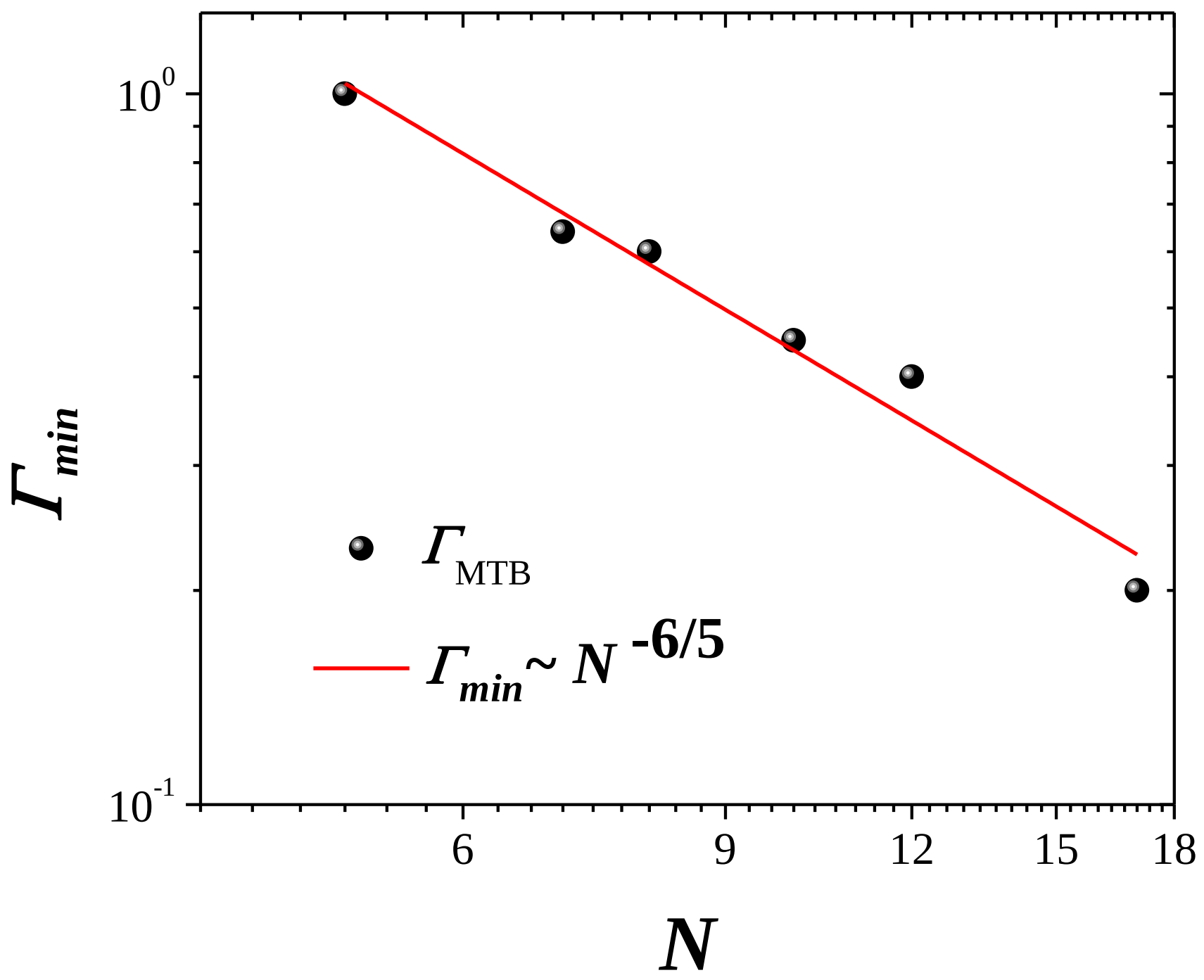

**Figure 8.** Grafting density at the minimum value of the friction coefficient ($\mu$), extracted from Figs. 2 and 3, $\Gamma_{min}$, as a function of the polymerization degree, $N$**.** The line is the fit to the scaling appropriate for the mushroom to brush transition (MBT), obtained when the scaling exponent of $N$ is -6/5, see eq. (2).

Finally, the friction coefficient's first minimum is about the same for all cases, when the value of $\phi$ is close to 0.5; that is, the minimum friction is obtained when the number of solvent particles is about twice the number of monomers making up the chains.

## IV CONCLUSIONS

The tribology of linear chains grafted to parallel surfaces in the regime close to the MBT under stationary flow is studied here, with particular emphasis on their scaling properties, using mesoscale numerical simulations. The friction coefficient is found to display highly non – monotonous behavior as a function of increasing chain grafting density, while the viscosity increases steadily with grafting density. The change in the friction coefficient with grafting density is found to be due to the transition of the collective interactions of the chains, from being isolated in mushroom – like fashion, to becoming dense polymer brushes. In the grafting density range where this mushroom – to – brush transition takes place, the viscosity and the friction coefficient obey scaling laws determined by the intrinsic characteristics of such transition, and by the quality of the solvent in which the chains are dissolved. This is the first reported work on the scaling of the tribology of polymer chains near the MBT, to the best of our knowledge. Most recent experimental and theoretical/numerical works have focused on the scaling of static properties of brushes at the MBT, with particular emphasis on the scaling of the brushes' height [30 – 33]. However, detailed knowledge of the scaling aspects of polymer chains near the MBT under flow is increasingly necessary not only to advance our basic understanding of soft condensed matter, but also for the optimized design of new nanomaterials with specifically tailored tribology.

## V. ACKNOWLEDGEMENTS

The authors would like to thank ABACUS, CONACyT grant EDOMEX-2011-C01-165873, for funding, and Claudio Pastorino (CAC, CNEA-CONICET) for educational discussions. AGG was funded in part by Proinnova – CONACYT, through grant 231810.

The authors declare no competing financial interest.

## VI. APPENDIX

In this Appendix we include the figures with full results on all cases modeled, which were shown for only a few cases in the main part of this article to keep it brief.

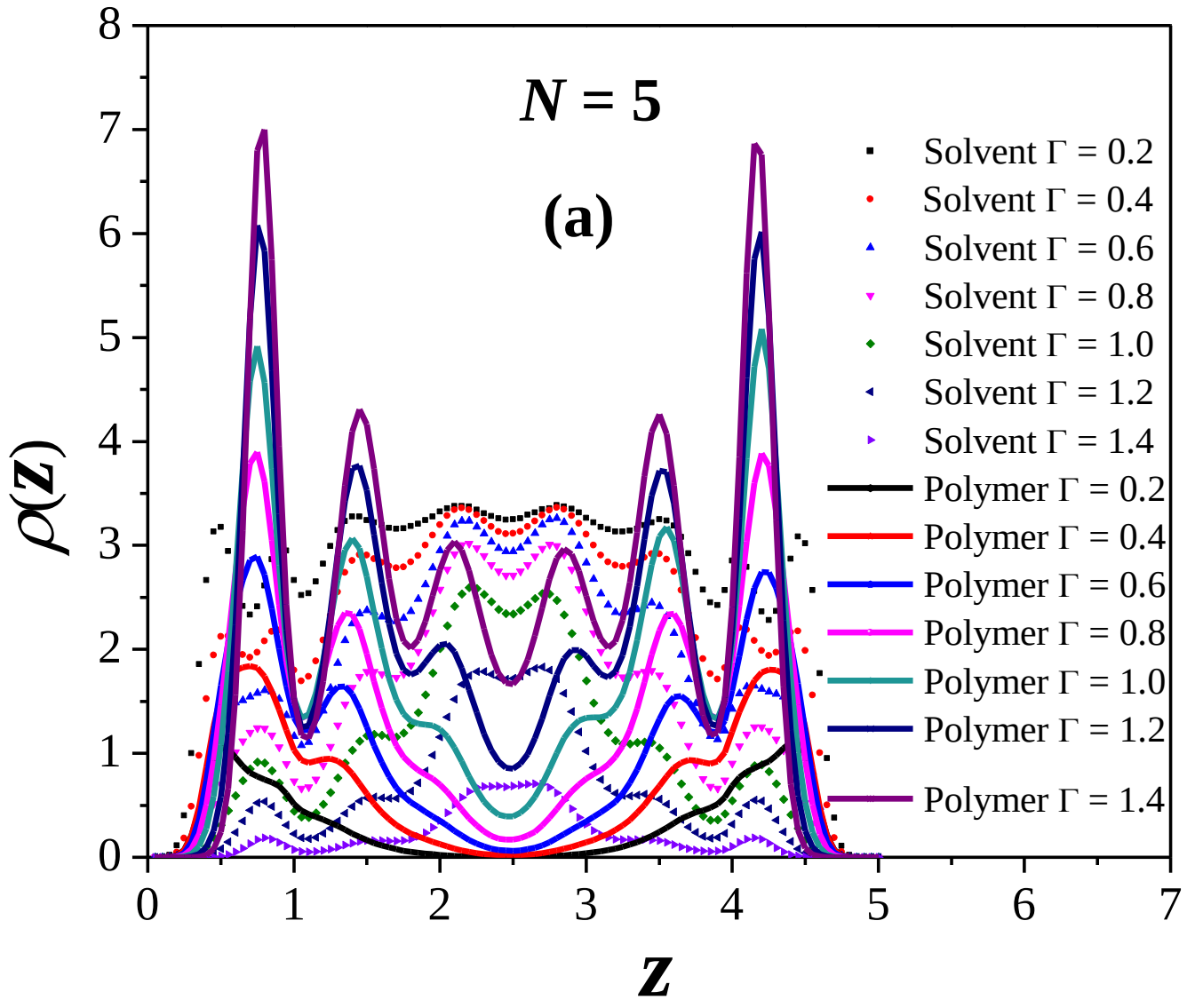

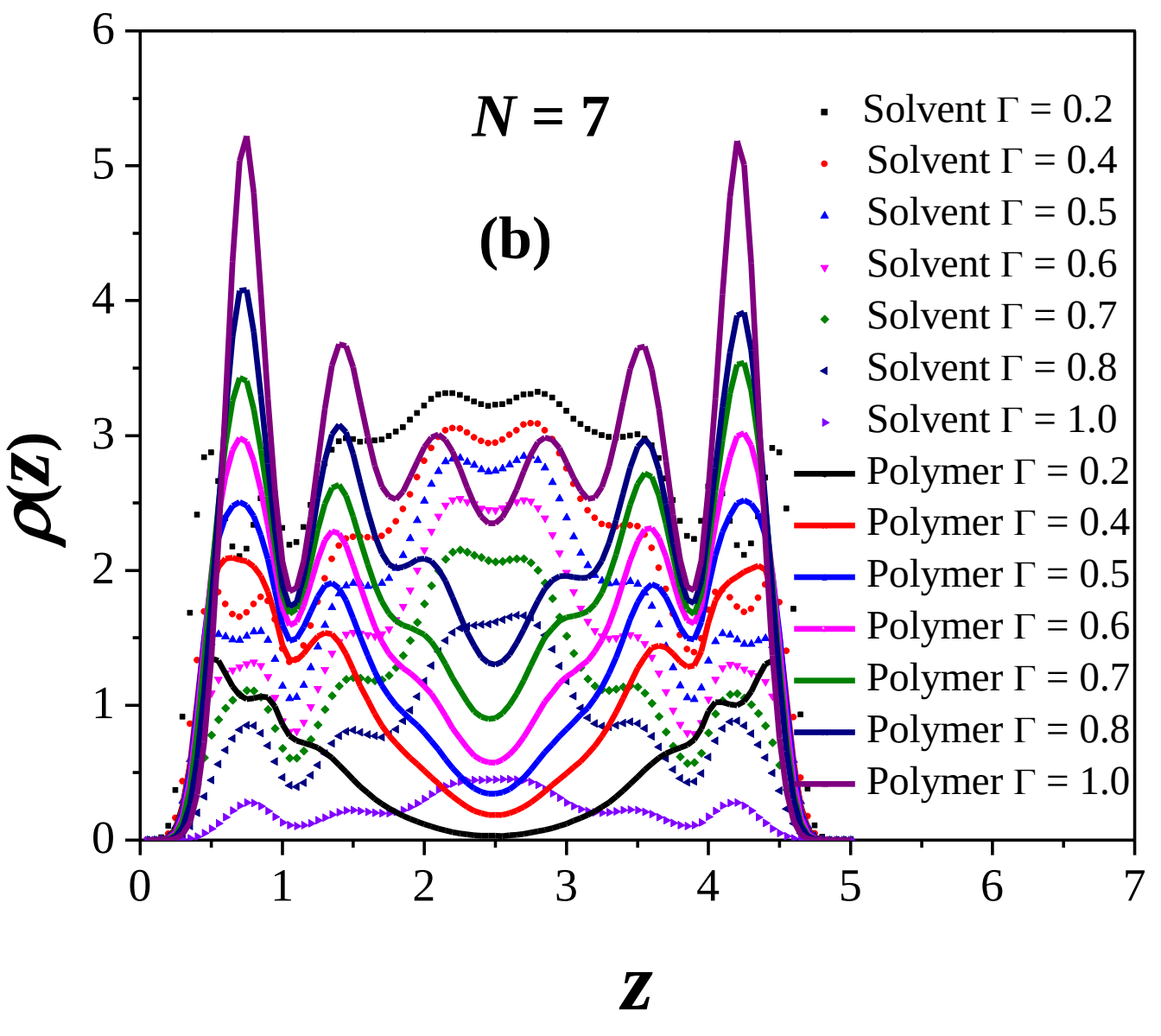
N = 7
(b)
Solvent Γ = 0.2
Solvent Γ = 0.4
Solvent Γ = 0.5
Solvent Γ = 0.6
Solvent Γ = 0.7
Solvent Γ = 0.8
Solvent Γ = 1.0
Polymer Γ = 0.2
Polymer Γ = 0.4
Polymer Γ = 0.5
Polymer Γ = 0.6
Polymer Γ = 0.7
Polymer Γ = 0.8
Polymer Γ = 1.0
ρ(z)
z

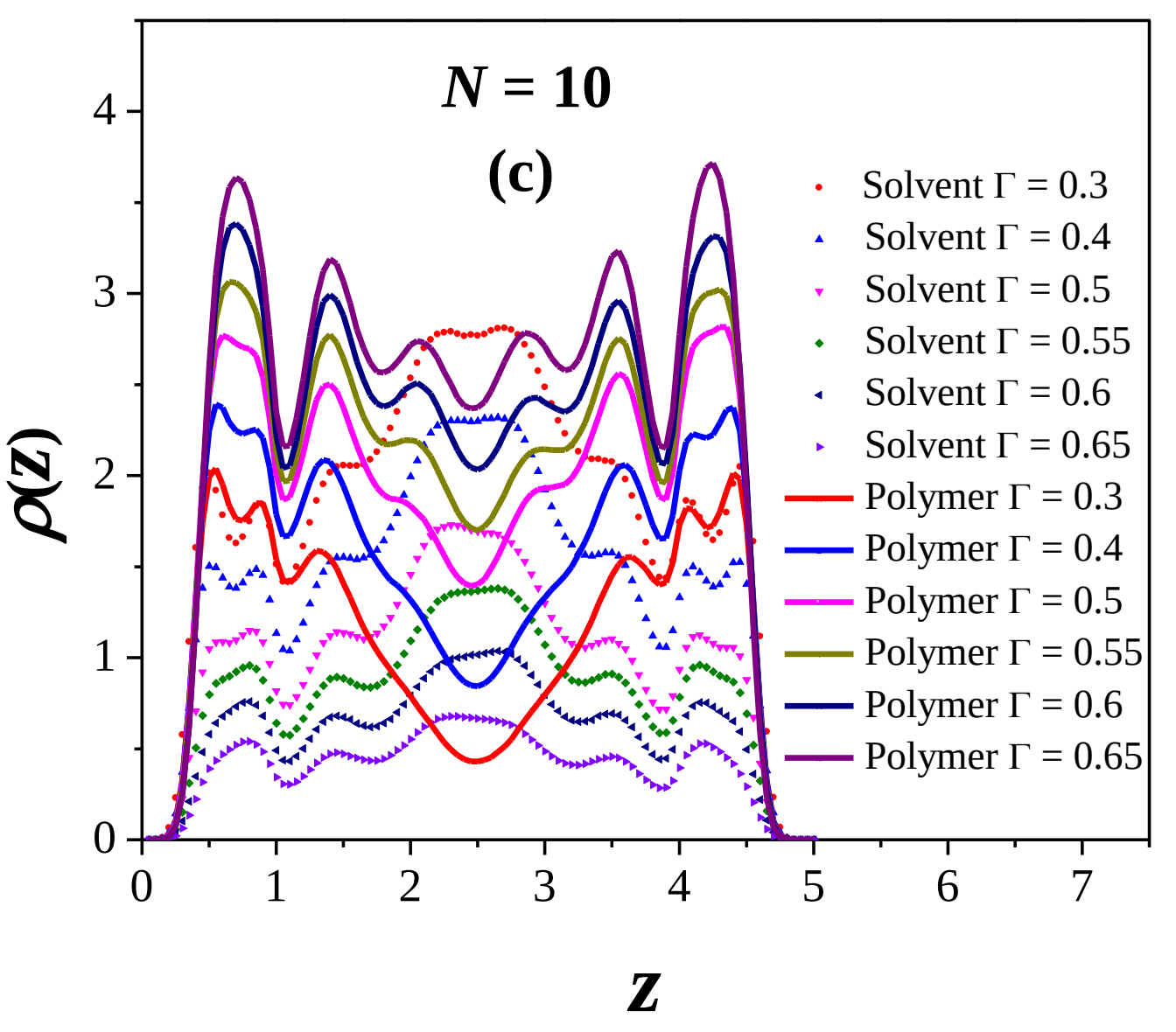
N = 10
(c)
Solvent Γ = 0.3
Solvent Γ = 0.4
Solvent Γ = 0.5
Solvent Γ = 0.55
Solvent Γ = 0.6
Solvent Γ = 0.65
Polymer Γ = 0.3
Polymer Γ = 0.4
Polymer Γ = 0.5
Polymer Γ = 0.55
Polymer Γ = 0.6
Polymer Γ = 0.65
ρ(z)
z

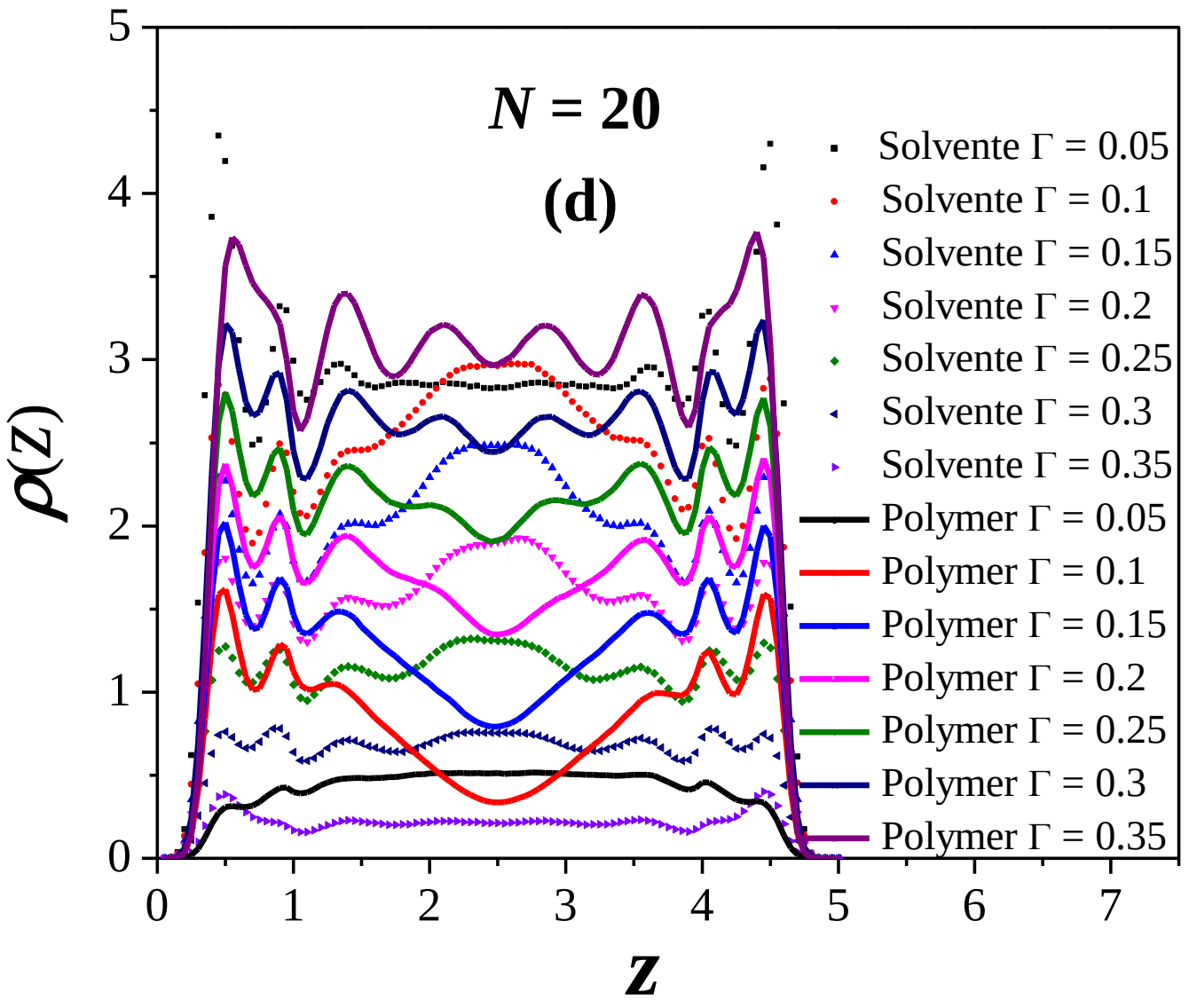

**Figure A1**. Density profiles for the solvent (symbols) and brush monomers (lines) of the larger system at increasing values of the grafting density Γ, for four values of the polymerization degree, namely *N* = 5 (a), 7 (b), 10 (c) and 20 (d), as indicated in each graph. All quantities are reported in DPD units.

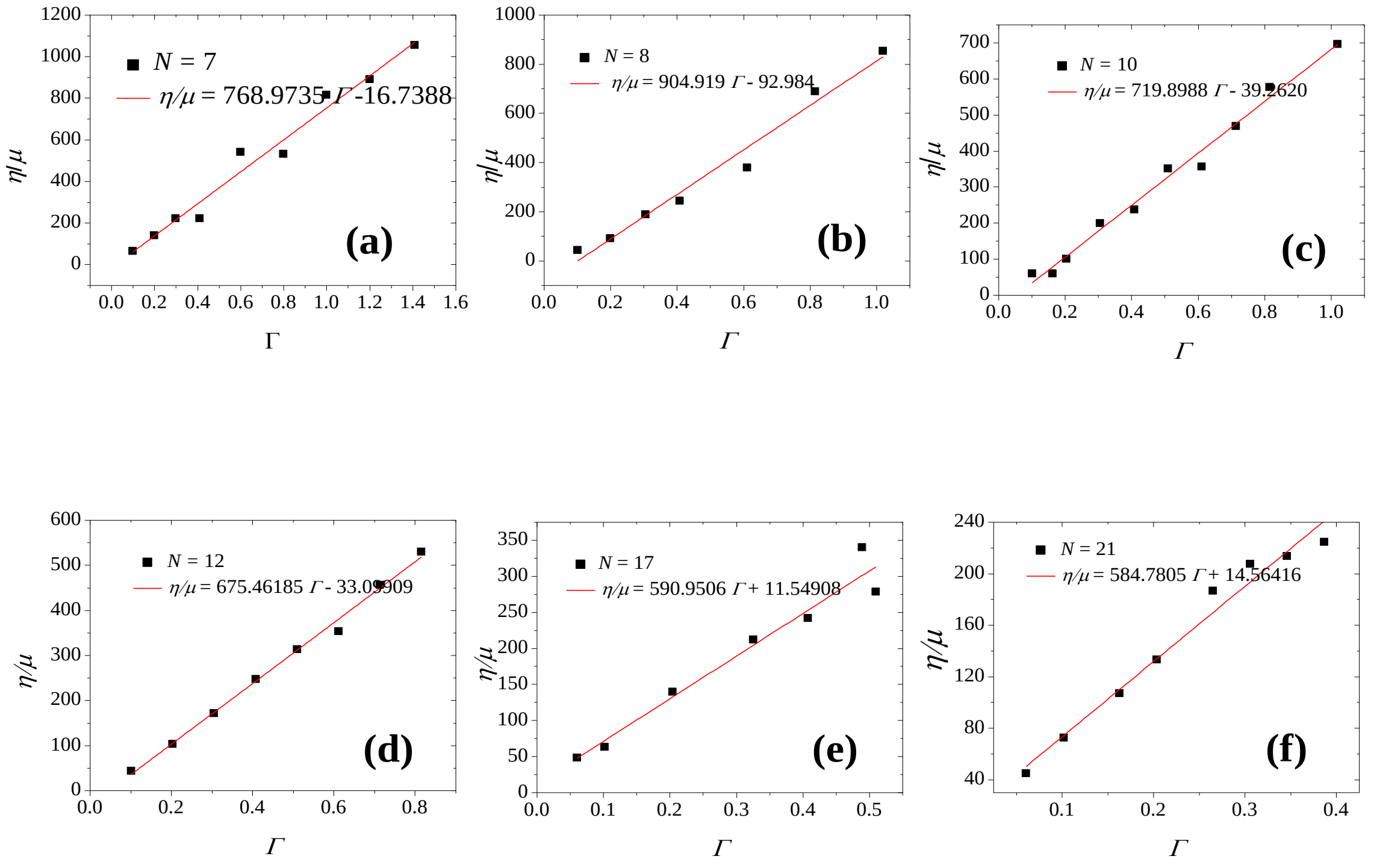

**Figure A2.** The extrapolation length $b$ as a function of the grafting density Γ for different values of $N$: (a) $N$ = 7, (b) $N$ = 8, (c) $N$ = 10, (d) $N$ = 12, (e) $N$ = 17 and (f) $N$ = 21. The lines are the best linear fits. All quantities are reported in reduced units.